\documentstyle[12pt,aaspp4,flushrt]{article}
\lefthead{}
\righthead{}
                            
\begin{document}

\title{A Determination of the Local Density of White Dwarf Stars}
                                
\author{J. B. Holberg\altaffilmark{1}}
\author{Terry D. Oswalt\altaffilmark{2}}
\author{E. M. Sion\altaffilmark{3}}

\altaffiltext{1}{Lunar and Planetary Laboratory, Gould-Simpson Bld., 
University of Arizona, Tucson, AZ 85721 holberg@argus.lpl.arizona.edu}
 
\altaffiltext{2}{Dept. Physics \& Space Sciences, Florida Institute of Technology, 150 W. University Boulevard,
Melbourne, FL 32901; oswalt@tycho.pss.fit.edu}

\altaffiltext{2}{And: National Science Foundation, Div. of Astronomical Sciences, 4201 Wilson Blvd, Arlington, VA 22230; toswalt@nsf.gov}

\altaffiltext{3}{Department of Astronomy and Astrophysics, Villanova University, Villanova PA 19085; emsion@ucis.vill.edu}

 \begin{abstract}
The most recent version of the Catalog of Spectroscopically Identified White Dwarfs lists 2249
white dwarf stars.  Among these stars are 118 white dwarfs that have either reliable trigonometric
parallaxes or color-based distance moduli which place them at a distance within 20 pc of the Sun. 
Most of these nearby white dwarfs are isolated stars, but 35 (30 \% of the sample) are in binary
systems, including such well known systems as Sirius A/B, and Procyon A/B.  There are also three
double degenerate systems in this sample of the local white dwarf population.  The sample of local
white dwarfs is largely complete out to 13 pc and the local density of white dwarf stars is found to
be 5.5 $\pm$ 0.8 x 10$^{-3}$  pc$^{-3}$ with a corresponding mass density of 
3.7$\pm$0.5 x10$^{-3}$ M$_{\odot}$  pc$^{-3}$.

 \end{abstract}


\section{INTRODUCTION}
There is presently considerable interest in the stellar and sub-stellar components of the volume 
surrounding the Sun.  The primary registry for the known stars within 25 pc of the
Sun  has been The Catalogue of Nearby Stars, 3$^{rd}$ Edition (Gliese \& Jahreise 1991).  Recently,
however, a comprehensive effort to compile information on virtually all stellar and sub-stellar sources
within 20 pc of the Sun has been jointly undertaken by NASA and NSF. This program, called 
NSTARS, which is an effort to support  future NASA missions such as the Space Interferometry Mission
(SIM) and the Terrestrial Planet Finder, also has the scientific goals of understanding the stellar
population near the Sun and its evolutionary history. This population overwhelming consists of low
luminosity stars which are difficult to study at great distances from the Sun.   A significant component
of this local stellar population of low luminosity stars are white dwarfs.  Such white
dwarf stars are currently of major interest for several reasons.  First, they represent a history of star
formation and stellar evolution in the Galactic plane and the luminosity function of these stars can be
used to place a lower limit on the age of the Galactic disk (Liebert, Dahn, Monet 1988, Oswalt et al.
1996).  Second, white dwarfs, in particular the cooler stars, have been suggested as the origin of
the MACHO lensing objects seen in lensing surveys (Kawaler 1996, Graff et al. 1998).  Third,
estimates of the local density of white dwarfs are  important to a full understanding of the mass
density of the Galactic plane (Bahcall 1984). 

All methods of obtaining estimates of the space density of white dwarfs begin with a well defined
observational sample and estimates of its completeness.  One method is to obtain a magnitude  limited
sample of white dwarfs obtained from large color surveys such as the Palomar-Green survey (Green,
Schmidt, \& Liebert 1986).   A second method relies on a proper motion limited sample and uses the 
1/V$_{max}$ procedure of Schmidt (1968) to correct for kinematic bias (Wood \& Oswalt 1998).  A third
possibility is to use a volume limited sample of very high completeness.

In this paper we use the 4$^{th}$ edition of the Catalog of Spectroscopically Identified White Dwarfs
(McCook \& Sion 1999, hereafter MS99) to identify the known white dwarfs within a sphere of radius
20 pc around the Sun.  We have chosen the distance of 20 pc because it corresponds to the volume
of the NSTARS database while also being a subset of the Catalogue of Nearby Stars.  As we shall
see, this distance also contains the spherical volume in which the sample of known white dwarfs
is reasonably complete.   MS99 contains 2249 white dwarfs and nearly doubles the number of known
degenerate stars compared to the previous version (McCook \& Sion 1987) of the catalog.  In
addition to new stars, it contains a great amount of new information on previously identified stars,
including spectral classification, updated photometric measurements, trigonometric parallaxes,
absolute magnitudes, etc.  The new MS99 catalog is therefore a valuable source for up-to-date
information on the white dwarfs residing near the Sun. 

In \S 2.0 we introduce the sample of the local population of white dwarfs.  In \S 3.0 we discuss the
distribution and completeness of this local sample.  We also estimate the space density and mass
density of white dwarfs and discuss the nature of the sample of white dwarfs near the Sun. A
preliminary version of this paper was presented at the NSTARS meeting in 1999.

\section{The Population of the Local White Dwarfs}
Using the search capabilities of the University of Arizona White Dwarf Database
\footnote{http://procyon.lpl.arizona.edu}, 
the MS99 catalog was searched for all degenerate stars having parallaxes $\pi$ $\geq$ 0.05$\arcsec$ 
and for stars with photometric distances corresponding to V-M$_{v}$ $\leq$ 1.505.  The list of stars 
satisfying these two criteria was then examined and obvious anomalies such as Feige 24, a spectroscopic 
binary with a composite DA + dMe spectrum, were eliminated.  The final list (hereafter, 
the {\bf {\em local sample}}) consisted of 118 degenerate stars, including three double degenerate 
systems.  This represents  approximately 6\% of
the total MS99 catalog and 5\% of the stars currently in the NSTARS database. The list of stars
contained in the local sample is given in Table 1 along with the MS99 spectral type, visual and
absolute magnitudes, trigonometric parallaxes, photometric and trigonometric parallax distances, and
the adopted distance for each star as well as an indication of the presence of any binary companions.
There are 116 enteries in Table 1, two of the double degenerate systems (WD 0727+482 and WD0135-052) 
are spectroscopic or unresolved visual binaries and are listed by system while the entry for WD0747+073, 
a resolved visual binary, contains both components.  In Table 1 stars which are included in MS99 
are designated by their WD number, while stars not in  MS99 are listed with alternate catalog names.

In establishing the final adopted distances in Table 1 we have considered both trigonometric
parallaxes and photometric parallaxes.  The photometric data 
provided in MS99, which attempts to report all available observations, is not
homogeneous and is far from uniform in quality.  We have therefore adopted the following scheme
for using this data to obtain photometric distances.  We use, wherever possible, the Johnson $B-V$,
Str\"{o}mgren $b-y$ and the multichannel $g-r$ colors to estimate absolute $V$ magnitudes from the color
magnitude relations employed by MS99.  For stars with multiple observations of these quantities, we
have used an average of the available colors unless one or more of the observations was obviously
discrepant or a redundant reporting of a prior observation.  Final absolute magnitudes, derived from
the different color magnitude relations were also averaged. Where there existed gross disagreement
among the photometric absolute magnitudes, the issue was resolved by looking at other data such as
effective temperatures and gravities or by the trigonometric parallax.   For the apparent visual
magnitudes, a similar averaging was used with the Johnson $V$ magnitude assumed to be on the same
scale as the Str\"{o}mgren $y$ magnitude.  The multichannel magnitudes were not included unless no other
sources of apparent magnitude were available.  For the trigonometric parallaxes the following sources
were used in order of preference; {\em Hipparcos} (ESA, 1997) values, Van Altena et al. (1997) values,
US Naval Observatory Parallax Program values or other values. In Fig. 1 we plot the trigonometric
distances against the photometric distances for those stars possessing both estimates. The standard 
deviation about the 1:1 correlation line in Fig. 1 is 2.9 pc.  This scatter is primarily due to the 
inherent uncertainty in the photometric distance estimates arising from color dependence due to 
stellar gravity and spectral type which is not contained in the color magnitude relations. 

\vspace{8cm}
\begin{figure}
       \plotfiddle{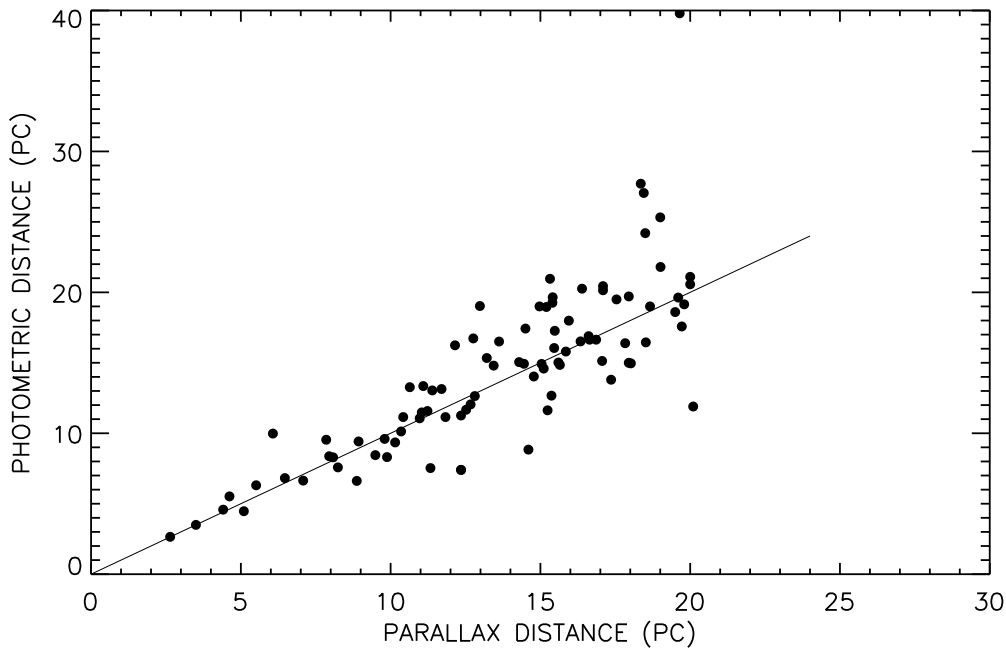}{1cm}{90}{60}{60}{220}{10}
       \caption{A comparison of the photometric and trigonometric distances for white dwarfs in the local sample.}
       \end{figure}
\vspace{5cm}

\makeatletter
\def\jnl@aj{AJ}

\ifx\revtex@jnl\jnl@aj\let\tablebreak=\nl\fi
\makeatother

\def\fnum@ptable{TABLE \thetable}
\def\fnum@ptablecont{TABLE \thetable---{\it Continued}}

\def\set@mkcaption{\long\def\@makecaption##1##2{
    \center\Large\sc##1\\[.5ex]##2\endcenter\vskip 1.5ex}}
\def\set@mkcaptioncont{\long\def\@makecaption##1##2{
    \center\Large\sc##1\endcenter\vskip 1.5ex}}

\begin{planotable} {lccccccccc} 
\tablenum{1}
\tablewidth{42pc}
\tablecaption{Known White Dwarfs within 20pc}
\vspace{3mm}
\tablehead{
\colhead{WD Number}    &
\colhead{Type}    &
\colhead{V}    &
\colhead{M$_{v}$}    &
\colhead{$\pi$(mas)}    &
\colhead{D$_{v}$(pc)}    &
\colhead{D$_{\pi}$(pc)}  &
\colhead{D$_{adp}$(pc)}  &
\colhead{System$^{1}$}}
\startdata
WD0000$-$345&  DC9&   14.94&     14.01&     75.7&   15.34&   13.21&         13.21 &      \nl
WD0009$+$501&  DA8&   14.37&     14.07&     90.6&   11.48&   11.04&         11.04 &      \nl
WD0011$-$134&  DC8&   15.88&     14.53&     51.3&   18.60&    19.5&          19.5 &      \nl
WD0034$-$211&  DA7&   14.43&     13.83&    .....&   11.69&    ....&          11.7 &    b \nl
WD0038$-$226&  DC9&   14.52&     14.92&    101.2&    8.31&    9.88&          9.88 &      \nl
WD0046$+$051&  DZ7&   12.39&     14.09&   226.95&    4.58&    4.41&          4.41 &      \nl
WD0115$+$159&  DQ6&   13.85&     12.43&     64.9&   19.26&    15.4&          15.4 &      \nl
WD0123$-$262&  DC7&   15.00&     13.08&      54.&    24.2&    18.5&          18.5 &      \nl
WD0135$-$052&  DA7&   12.83&     13.49&      81.&    7.40&   12.35&         12.35 &    dd \nl
WD0141$-$675&  DA7&   13.87&     13.96&     102.&    9.60&     9.8&           9.8 &      \nl
WD0148$+$467&DA3.5&   12.44&     11.45&    63.08&   15.80&   15.85&         15.85 &      \nl
WD0148$+$641&  DA6&   14.00&     12.80&    .....&    17.5&    ....&          17.5 &    b \nl
WD0208$+$396&  DA7&   14.52&     13.42&     59.3&   16.64&   16.86&         16.86 &      \nl
WD0213$+$427&  DA9&   16.21&     14.75&     51.0&   19.63&    19.6&          19.6 &      \nl
WD0230$-$144&   DA&   15.76&     14.88&     64.0&   15.02&    15.6&          15.6 &      \nl
WD0235$+$064&  DA6&   15.09&     13.96&    .....&   16.84&   .....&         16.84 &      \nl
WD0245$+$541&  DA9&   15.50&     15.47&     96.6&   10.13&   10.35&         10.35 &      \nl
WD0310$-$688&  DA3&   11.39&     11.54&     98.5&    9.35&   10.15&         10.15 &      \nl
WD0311$-$543&  DZ7&   14.83&     14.41&    .....&   12.15&    ....&         12.15 &      \nl
WD0322$-$019& DA10&   16.22&     14.90&    .....&   18.36&    ....&         18.36 &      \nl
WD0326$-$273&  DA5&   13.56&     12.86&     57.6&    13.8&   17.36&         17.36 &    b \nl
WD0341$+$182&  DQ8&   15.21&     13.52&     52.6&    21.8&   19.01&         19.01 &      \nl
WD0357$+$081&  DC9&   15.89&     14.82&     56.1&   16.39&   17.83&         17.83 &      \nl
WD0413$-$077&  DA3&    9.52&     11.27&   196.24&    4.47&    5.10&          5.10 &    b \nl
WD0419$-$487&  DA8&   14.37&     14.80&     ....&    11.0&    ....&          11.0 &    b \nl
WD0423$+$120&  DC8&   15.41&     14.37&     ....&   16.39&    ....&         16.39 &      \nl
WD0426$+$588&  DC5&   12.44&     13.43&   181.36&    6.31&    5.51&          5.51 &    b \nl
WD0433$+$270&  DC8&   15.82&     14.68&     60.2&   16.89&   16.61&         16.61 &    b \nl
WD0435$-$088&  DQ7&   13.77&     14.14&    105.4&    8.45&    9.49&          9.49 &      \nl
WD0509$+$168&   DA&   13.35&     14.70&    .....&     8.5&   .....&           8.5 &      \nl
WD0532$+$414&  DA7&   14.76&     13.44&    .....&   18.37&   .....&         18.37 &      \nl
WD0548$-$001&  DQ9&   14.58&     13.95&     90.2&   13.35&   11.09&         11.09 &      \nl
WD0552$-$041& DZ11&   14.46&     15.29&    154.6&    6.82&    6.47&          6.47 &      \nl
WD0553$+$053&  DA9&   14.11&     14.50&    125.0&    8.37&    7.95&          7.95 &      \nl
WD0628$-$020&   DA&    15.3&     15.13&    .....&   10.81&    ....&         10.81 &    b \nl
WD0642$-$166&  DA2&     8.3&     11.18&   379.21&    2.65&    2.64&          2.64 &    b \nl
WD0644$+$025&  DA8&   15.70&     13.54&     54.2&   27.05&   18.45&         18.45 &      \nl
WD0644$+$375&  DA2&   12.09&     10.62&    64.91&   19.65&   15.41&         15.41 &      \nl
WD0657$+$320&  DC9&   16.62&     15.23&     53.5&   19.00&   18.66&         18.66 &      \nl
WD0659$-$063&  DA8&   15.42&     15.16&     81.0&   11.26&   12.35&         12.35 &      \nl
WD0727$+$482&  DC9&   14.65&     15.27&     88.3&    7.53&   11.33&         11.33 &    dd \nl
WD0728$+$642&  DC9&   16.38&     15.08&     ....&   18.23&    ....&         18.23 &      \nl
WD0736$+$053&  DA4&   10.92&     13.20&    285.9&    3.50&    3.50&          3.50 &    b \nl
WD0738$-$172&  DZ6&   13.02&     13.15&      112&    9.42&    8.93&          8.93 &    b \nl
WD0743$-$336&  DC9&   16.59&     15.20&    65.79&   18.96&   15.20&         15.20 &    b \nl
WD0747$+$073.1&DC9&   16.98&     15.43&     58.5&   20.44&   17.09&         17.09 &    dd \nl
WD0747$+$073.2&DC9&   16.98&     15.46&     58.5&   20.15&   17.09&         17.09 &    dd \nl
WD0752$-$676&  DQ9&   14.08&     14.97&    141.2&    6.64&    7.08&          7.08 &      \nl
WD0824$+$288&   DA&   14.22&     13.87&    .....&   11.76&   .....&         11.76 &    b \nl
WD0839$-$327&  DA6&   11.88&     12.78&    112.7&    6.62&    8.87&          8.87 &      \nl
WD0912$+$536&  DC7&   13.87&     13.63&    96.  &   11.15&   10.42&         10.42 &      \nl
WD0939$+$071&  DA2&   14.90&     13.52&    .....&   18.88&   .... &         18.88 &      \nl
WD1013$-$559&  DZ9&   15.09&     14.78&    .....&   11.54&   .... &         11.54 &      \nl
WD1019$+$637&  DA7&   14.70&     13.61&    61.2 &   16.52&   16.34&         16.34 &      \nl
WD1033$+$714&  DC9&   16.89&     15.27&    50.  &   21.1 &   20   &            20 &      \nl
WD1036$-$204&  DQ9&   16.28&     15.67&    94.  &   13.27&   10.64&         10.64 &      \nl
WD1043$-$188&  DQ9&   15.5 &     14.38&    78.4 &   16.73&   12.76&         12.76 &    b \nl
WD1055$-$072&  DA7&   14.32&     13.27&    82.3 &   16.24&   12.15&         12.15 &      \nl
WD1121$+$216&  DA7&   14.24&     13.39&    74.4 &   14.80&   13.44&         13.44 &      \nl
WD1126$+$185&  DC8&   13.79&     14.08&    .....&    8.77&   .....&          8.77 &      \nl
VB 4      &  DC &   15.  &     .....&    104.5&   .....&   9.57 &          9.57 &    b \nl
WD1134$+$300&  DA2&   12.45&     10.84&    65.28&   20.96&   15.32&         15.32 &      \nl
WD1142$-$645&  DQ6&   11.48&     12.77&    216.4&   5.52 &   4.62 &          4.62 &      \nl
WD1223$-$659&  DA &   13.93&     13.67&    .....&   10.79&   .... &         10.79 &      \nl
WD1236$-$495&  DA6&   13.83&     12.30&    61.0 &   20.26&   16.39&         16.39 &      \nl
WD1257$+$037&  DC9&   15.83&     14.80&    60.3 &   16.05&   15.46&         15.46 &      \nl
WD1309$+$853&  DC9&   15.98&     15.09&    70.0 &   15.05&   14.29&         14.29 &      \nl
WD1310$-$472&  DC9&   17.11&     16.24&    66.5 &   14.92&   15.04&         15.04 &      \nl
WD1327$-$083&DA3.5&   12.32&     11.45&    55.5 &   14.96&   18.02&         18.02 &    b \nl
WD1334$+$039&  DZ9&   14.66&     15.26&    121.4&   7.58 &   8.24 &          8.24 &      \nl
WD1344$+$106&  DA7&   15.10&     13.53&    49.9 &   20.57&   20.  &            20 &      \nl
WD1345$+$238&  DC9&   15.65&     15.41&    84.5 &   11.15&   11.83&         11.83 &    b \nl
WD1444$-$174&  DC8&   16.46&     15.25&    69.0 &   17.43&   14.5 &          14.5 &      \nl
WD1514$+$033&  DA &   14.02&     12.57&    .....&   19.50&   .....&         19.50 &      \nl
WD1544$-$377&  DA7&   12.78&     12.45&    65.6 &   11.63&   15.24&         15.24 &    b \nl
WD1609$+$135&  DA6&   15.10&     12.89&    54.5 &   27.71&   18.35&         18.35 &      \nl
WD1620$-$391&  DA2&   11.00&     10.49&    78.04&   12.64&   12.81&         12.81 &    b \nl
WD1626$+$368&  DZ5.5& 13.84&     12.57&    62.7 &   17.99&   15.95&         15.95 &      \nl
WD1633$+$433&  DA8&   14.83&     14.01&    66.2 &   14.59&   15.11&         15.11 &      \nl
WD1633$+$572&  DQ8&   15.00&     14.13&    69.2 &   14.93&   14.45&         14.45 &    b \nl
BD$+$76$^{\circ}$614B& DA&13&     10&     50.9&    39.8&   19.65&          19.5 &    b \nl
WD1647$+$591&  DA4&   12.23&     12.01&    91.13&   11.06&   10.97&         10.97 &      \nl
WD1705$+$030&  DZ7&   15.19&     13.74&    57.0 &   19.50&   17.54&         17.54 &      \nl
WD1717$-$345&  DA &   16.38&     15.20&    .....&   17.26&   .....&         17.26 &      \nl
WD1743$-$132&  DA7&   14.24&     13.16&    54.  &   16.45&   18.52&         18.52 &    b \nl
WD1748$+$708&  DQ8&   14.15&     14.16&    164.7&   9.98 &   6.07 &          6.07 &      \nl
WD1756$+$827&  DA7&   14.32&     13.46&    63.9 &   14.85&   15.65&         15.65 &      \nl
WD1820$+$609&  DC9&   15.65&     15.25&    78.9 &   12.05&   12.67&         12.67 &      \nl
WD1829$+$547&  DQ7&   15.50&     14.11&    66.8 &   19.00&   14.97&         14.97 &      \nl
WD1900$+$705&  DA4.5& 13.22&     11.82&    77.0 &   19.03&   12.98&         12.98 &      \nl
WD1917$+$386&  DC7&   14.60&     14.01&    85.5 &   13.14&   11.70&         11.70 &      \nl
WD1917$-$077&  DBQA5& 12.29&     11.97&    89.08&   11.58&   11.23&         11.23 &    b \nl
WD1919$+$145&  DA5&   13.00&     11.59&    50.5 &   19.15&   19.80&         19.80 &      \nl
WD1935$+$276&  DA4.5& 12.96&     12.08&    55.7 &   15.00&   17.95&         17.95 &      \nl
WD1953$-$011&  DA6&   13.69&     13.11&    87.8 &   13.04&   11.39&         11.39 &      \nl
WD2002$-$110&  DC9&   16.89&     15.42&    55.7 &   19.71&   17.95&         17.95 &      \nl
WD2007$-$219&  DA6&   14.40&     13.10&    .....&   18.22&   .... &         18.22 &      \nl
WD2007$-$303&  DA4&   12.18&     11.67&    65.06&   12.67&   15.37&         15.37 &      \nl
WD2032$+$248&  DA2.5& 11.52&     10.79&    67.65&   14.03&   14.78&         14.78 &      \nl
WD2047$+$372&  DA4&   12.96&     11.67&    .... &   18.16&   .... &         18.16 &      \nl
WD2048$+$263&  DC9&   15.60&     15.22&    49.8 &   11.90&   20.1 &          15.5 &      \nl
WD2054$-$050&  DC9&   16.64&     15.45&    64.6 &   17.27&   15.48&         15.48 &    b \nl
WD2055$+$221&  DC &   .... &     .... &    73.24&    ....&   13.65&         13.65 &      \nl
WD2105$-$820&  DA6&   13.57&     12.67&    58.6 &   15.13&   17.06&         17.06 &      \nl
WD2117$+$539&  DA3.5& 12.35&     11.12&    50.7 &   17.58&   19.72&         19.72 &      \nl
WD2133$+$463&  DC &   17.8 &     16.77&    .... &   16.07&   .... &         16.07 &    b \nl
WD2140$+$207&  DQ6&   13.24&     12.90&    79.9 &   11.67&   12.52&         12.52 &      \nl
WD2154$-$512&  DQ7&   14.74&     15.01&    68.5 &   8.84 &   14.6 &          14.6 &    b \nl
WD2246$+$223&  DA5&   14.35&     12.33&    52.5 &   25.32&   19.0 &          19.0 &      \nl
WD2249$-$105&  DC11.5&17.45&     16.53&    .....&   15.26&   .....&         15.26 &    b \nl
WD2251$-$070&  DC13&  15.66&     16.07&    123.7&   8.29 &   8.08 &          8.08 &      \nl
WD2326$+$049&  DA4&   13.05&     11.96&    73.4 &   16.51&   13.62&         13.62 &      \nl
WD2341$+$322&  DA4&   12.93&     11.82&    60.11&   16.64&   16.64&         16.64 &    b \nl
WD2351$-$335&  DA5.5& 13.56&     13.95&    .....&   12.41&   .....&         12.41 &    b \nl
GD 1212   &  DA &   13.26&     12.29&    .....&   15.6 &   .....&          15.6 &      \nl
WD2359$-$434&  DA5&   12.91&     13.01&    127.4&   9.54 &   7.85 &          7.85 &    
\enddata 
\tablenotetext{1}{b = binary system, dd = double degenerate system}
\end{planotable}

\clearpage
\begin{center}
Notes to Table 1
\end{center}

\noindent{\em WD 0135-052  (L 870-2)} {\bf -}. A spectroscopic binary, consisting of a pair of DA stars with an
orbital period of 1.55 days (Saffer, Liebert, \& Olszewski 1988).

\noindent{\em WD 0413-077 (40 Eri B)} {\bf -}. A well studied nearby DA white dwarf.  The mass (0.501 M$_{\odot}$) is taken
from Provencal et al. (1998).

\noindent{\em WD 0419-487 (RR Cae)} {\bf -}. This is an eclipsing pre-cataclysmic binary system (DA6 + dM6) with
a 7.29 hr. period.  The distance (11pc) and mass (0.467 M$_{\odot}$) are taken from Bruch (1999).

\noindent{\em WD 0426+588 (Stein 2051B)} {\bf -}. A DC + dM binary system.

\noindent{\em WD 0642-166 (Sirius B)} {\bf -}. This is the nearest white dwarf as well as the hottest (T$_{eff}$ = 24,790 K)
and second  most massive (1.034 M$_{\odot}$, Holberg et al. 1998) of the white dwarfs in the local sample.

\noindent{\em WD 0727+482 (G107-70)} {\bf -}. A visual binary consisting of two DC white dwarfs (Sion et al. 1991)

\noindent{\em VB 4} {\bf -}.  Not in MS99, the {\em Hipparcos} parallax of the K0 V companion (HIP 56452) is 104.5 mas.

\noindent{\em WD 0736+053 (Procyon B)} {\bf -}. A difficult to observe companion to the F 5 IV-V star Procyon.  We
have adopted the {\em Hipparcos} parallax and  recent mass determination of M =  0.602 M$_{\odot}$ of Girard et
al. (2000).

\noindent{\em WD 0747+073} {\bf -}. A visual binary composed of two DC white dwarfs (Sion et al. 1991).

\noindent{\em WD 1620-391 (CD 38$^{\circ}$ 10980)} {\bf -}. A bright DA white dwarf in a wide binary system with a G5
V star.

\noindent{\em BD +76$^{\circ}$ 614B} {\bf -}. Not in MS99, the {\em Hipparcos} parallax (HIP 81139) of the K 7 companion is  50.9
$\pm$13.3 mas

\noindent{\em WD 1748+708 (G240-72))} {\bf -}.  This star is listed in MS99 as DXP9, however, BRL designate this
as a 'C$_{2}$H star'.  We have designated it as DQ8 using the BRL temperature.

\noindent{\em WD 1917-077 (LDS 678A)} {\bf -}.  This star is classified as a DBQA5 showing weak He I $\lambda$ 4471 and
C I features (Wesemael et al. 1993).  A gravitational redshift mass of 0.55 M$_{\odot}$  is given by Oswalt et
al. (1991).

\noindent{\em WD 2048+263 (LHS 3584)} {\bf -}. BRL find a low spectroscopic mass of (0.23  M$_{\odot}$)  and note the
possibility that this is a double degenerate system.

\noindent{\em GD 1212} {\bf -}.  Not in MS99, the photometric distance is from Gliese \& Jahriess (1991)

\begin{center}Not in Local Sample
\end{center}
There are 14 object present in TT99 which are not included in our local sample.

\noindent{\em WD 1246+586 \& WD1424+240} {\bf -}. Although listed in MS99, these two objects have been identified
as BL Lac objects ( see Fleming 1993 and Putney 1997).

\noindent{\em GD 806} {\bf -}. Not in MS 99, this star has been spectroscopically identified as a cool metal-rich
subdwarf. (I. Bues, private communication.)

\noindent{\em WD0713+584 (GD 294)} {\bf -}. This star was present in our original list of local sample candidates.  It
is also contained in Gliese \&  Jahreiss (1991) with an estimated distance of 12 pc.  However, it has
a negative {\em Hipparcos} parallax (HIP 35307, $\pi$ = -1.80 $\pm$ 2.97 mas) and Vauclair et al. (1997) classify
it as a possible sdB. Consequently, we have not included this star in our local sample.

\noindent{\em WD1026+002} {\bf -}.  A DA3 + dM4e system (Saffer et al. 1993).  Although Gliese \& Jahriess (1991)
estimate a photometric distance of 18 pc, the effective temperature of the DA is not consistent with
a distance less than about 37 pc.

\noindent{\em WD 1208+576, WD 1247+550, WD 1639+537, WD 2011+065, \& WD 2151-0156} {\bf -}.  All have
trigonometric parallaxes smaller than 0.05\arcsec (MS99). 

\noindent{\em WD 1655+210, WD 1821-131, WD 1840-111, WD, \& WD 2151-015} {\bf -}.  The photometric
distances place these stars outside the volume of the local sample.

\begin{center}Local Stars not in Tat \& Terzian
\end{center}
There are seven stars which we have included in our local sample, which are not in TT99.

\noindent{\em WD 1033+714 \& WD 1743-132} {\bf -}.  The trigonometric parallaxes of these stars place them within
the volume of the  local sample.

\noindent{\em WD 0509+168, WD 0628-020, WD 0824+288, WD 1717-345, \& WD 2133+463} {\bf -}.  The
photometric distances place these stars within the volume of the local sample.

\noindent{\em WD 2249-105, WD2351-335}{\bf -}.  White dwarfs in wide binary systems from Oswalt et al. (1996) and
Oswalt (private communication), respectively.

A prior survey of the population of local white dwarfs (Jahreiss 1987) estimated only 96 degenerate
stars within 20 pc of the Sun.  Even with the 23\% growth in the number of known white dwarfs in
the local sample, many of the conclusions reached by Jahreiss remain valid.  A similar determination
of the local sample has recently been made by Tat \& Terzian (1999, hereafter TT99) who compiled
a list of  white dwarfs within 20 pc, drawn principally from the prior version of the Catalog of
Spectroscopically Identified White Dwarfs (McCook \& Sion 1987) and the Catalogue of Nearby
Stars (Gliese \& Jarhiese 1991).  These authors, who where interested primarily in determining the
distribution of ionization in the local interstellar medium, found 121 white dwarfs within 20 pc of the
Sun.  In \S 3 we discuss individual stars of particular interest (see notes to Table 1) as well as 
the individual differences between our local sample and the TT99 sample.
   
\section{The Distribution and Completeness of the Local Sample}
In Fig. 2 we show the distribution in celestial coordinates of the local sample on an equal area
Hammer-Aitoff projection.  There are no obvious zones with low densities of stars.  There is,
however, an apparent $\sim$5:4 north-south asymmetry in the local sample. On the other hand, a
calculation of the centroid of the local sample shows that it is displaced from the Sun by a 
distance of only 1.4 pc in the direction of  $\alpha$ = 12.8$^{\circ}$ and $\delta$ = +0.5$^{\circ}$.  
This displacement should be compared with the expected standard deviation in the displacement of 118 
uniformly distributed stars within 20 pc of the Sun of 1.4 pc.  Alternately, the probability of 
finding hemispherical asymmetries in a binomial distribution of 118 objects which exceed 5:4 or 
are less than 4:5 is 0.23, demonstrating that the asymmetry should not be considered significant.  

\begin{figure}
       \plotfiddle{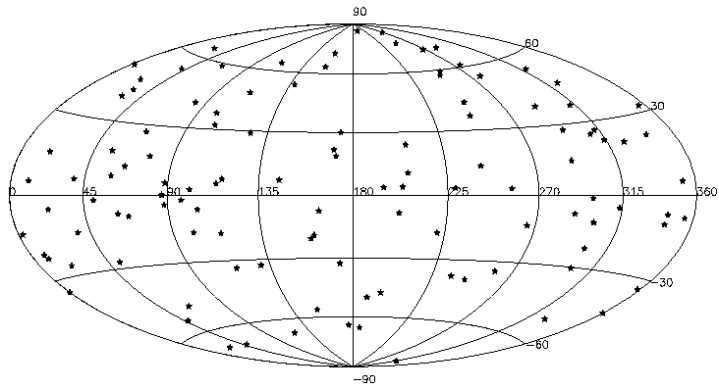}{4cm}{90}{40}{40}{135}{-10}
       \caption{A Hammer Aitoff equal area projection of the equatorial coordinates of the local sample of
white dwarfs within a radius of 20 pc of the Sun.}
       \end{figure}
\pagebreak
We discuss here the completeness of the local sample, in particular the completeness of the local 
sample out to a distance of 13 pc.  First, we note the distance of
13 pc is close enough that we are not excluding many faint cool white dwarfs based on apparent
brightness.  Due to the finite age of white dwarfs, the white dwarf luminosity function is truncated
near absolute magnitudes of M$_{v}$ = 16.2 so that at 13 pc the apparent magnitude of such stars are
brighter than 17.  Second, in Fig. 3 we have plotted a cumulative log $\sum$ N - log (distance) 
distribution
of the local sample.  Also shown in Fig. 3 is a line representing the expected number of white dwarfs
assuming a constant local density of these stars. As can be seen, this assumption appears valid out to
13 pc, at which point the observed number falls below the expected number;  as anticipated if the
local sample is incomplete beyond this distance.  The completeness of the white dwarfs out to 13 pc
is consistent with the earlier results of Jahreiss (1987), who also considered the number of white
dwarfs within 20 pc of the Sun from the 3$^{rd}$ Catalogue of Nearby Stars, and by Gleise, Jahreiss, \&
Upgren (1986) who came to the same conclusion for stars of all types in the 3$^{rd}$ Catalogue of Nearby 
Stars. Also Dawson (1986) finds that the LHS proper motion catalog (Luyten 1976) is complete to 
0.5$\arcsec$ yr$^{-1}$, which corresponds to a distance of $\sim$13 pc.  More recently, Fleming (1998) 
also finds that the sample of known M dwarfs is also largely complete to within 13 pc. Recently Flynn 
et al. (2000) have suggested a relatively low completeness for the LHS, however, these claims are 
contradicted by the results of Monet et al. (2000) who specifically searched for high proper motion 
stars over looked by the LHS.  From the relatively few such stars found, they confirm a relatively high 
completeness ($\sim$90 \%) for the LHS.  Our subsequent determination of the local space and mass density of 
white dwarfs is derived exclusively from our 13 pc sub-sample.  

We can also estimate the overall completeness of the local sample.  As noted above, the local sample 
appears complete out to 13 pc. Within this volume we find a total of 51 white dwarfs.  If we extrapolate
the corresponding space density to a distance of 20 pc there ought to be  approximately 186 white dwarfs 
within a distance of 20 pc compared to our observed  number of 118, thus the local sample is 
approximately 63\% complete out to 20 pc.  From this it can be anticipated that another 60 to 70 
white dwarfs remain to be discovered within 20 pc. The majority of these new stars will lie beyond 
13 pc and  be cool white dwarfs with visual magnitudes of 16 and 17; about 30\% will be in binary systems.

\begin{figure}
       \vspace{3in}
       \plotfiddle{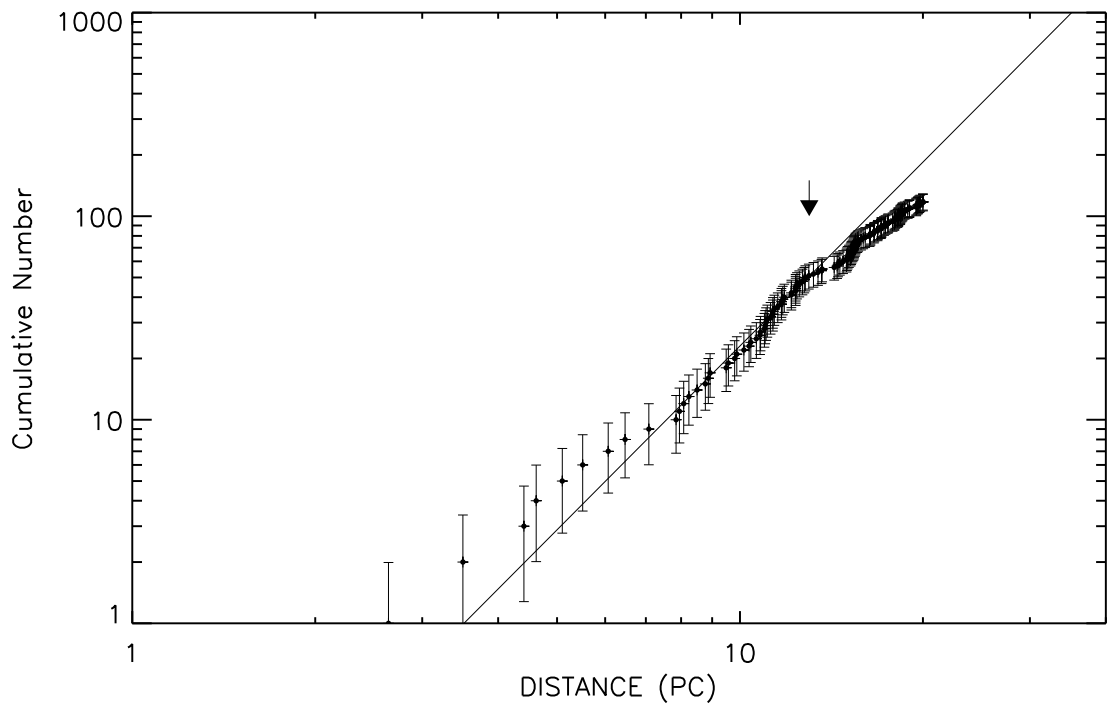}{1cm}{90}{65}{65}{250}{1}
       \caption{A cumulative Log $\sum$ N - Log(distance) plot of the white dwarfs in the local sample.  
The error bars are Poisson uncertainties in the cumulative number of stars. The straight line represents 
the expected number of white dwarfs having a mean space density of 5.5x10$^{-3}$  pc$^{-3}$.  The observed 
distribution of stars is consistent with a constant space density out to a distance of 13 pc (noted by arrow). 
The apparent excess at distances below $\sim$ 7 pc is due to small numbers of stars.  For example, removing 
Sirius B, the nearest white dwarf from the sample significantly reduces this feature.  Also note that 
the apparent ``bump'' $\sim$ 13 pc is near the proper motion cutoff of 0.5$\arcsec$ yr$^{-1}$ of the 
LHS Catalogue (Luyten 1976)}
\end{figure}

Most of the 51 stars found within 13 pc have trigonometric parallaxes, however, 10 stars possess only 
photometric distance estimates.  The question arises, what is the potential effect on the number of 
stars within 13 pc by including these photometric distances?  We have estimated this effect by using a
Monte Carlo calculation in which the photometric distances are replaced by a random variable having 
a mean equal to the photometric distance and a standard deviation equal to 2.9 pc, which we obtain 
from the correlation between trigonometric and photometric distance in Fig. 1.  The results of this 
calculation for 1000 trials is that the mean number stars with photometric distances less than 13 pc 
is 8.6 with a standard deviation of 2.5.  Thus, the 10 stars with photometric distances less than 13 
pc is consistent with this result.

\subsection{The Local Density of White Dwarfs}
Using the local sample we can directly estimate the local space density of white dwarfs.  Assuming
that the local sample of white dwarfs is complete out to a distance of 13 pc, we find a total of 51 
white dwarfs. The local density of white dwarfs represented by the straight line in Fig. 3 corresponds 
to 5.5x10$^{-3}$pc$^{-3}$.  If we assume that the local sample is representative of the mean density 
of white dwarfs in the Galactic plane near the Sun, then the uncertainty in this value can be 
determined from the Poisson variance of the number of stars within our 13 pc volume. Assuming this 
estimate to be representative of the Galactic plane near the Sun,  the mean  number density becomes 
5.5$\pm$0.8x10$^{-3}$ pc$^{-3}$. The mass density due to white dwarfs can also be directly 
determined.   Approximately 75\% of the stars in the local sample have published spectroscopic mass 
estimates.  These estimates are contained primarily in Bergeron, Ruiz, \& Leggett (1997, hereafter 
BRL), Leggett, Ruiz, \& Bergeron (1998) and several other references. The average mass of the local 
sample stars with published masses is 0.64 M$_{\odot}$, somewhat larger than the value of 
$\sim$0.57 M$_{\odot}$  found from white dwarf mass distributions derived from spectroscopic studies 
of the general population of white dwarfs (see Finley et al. 1997 and references therein).  However, 
the average mass of the local sample is near the value of  0.67 M$_{\odot}$ found by BRL for their 
sample of cool white dwarfs.  Among the effects BRL cite for this larger mass 
is the large fraction of non-DA stars in their sample.  It is also near the mean mass of 0.68 M$_{\odot}$ 
obtained by Silvestri et al. (2001) for the sample of common proper motion visual binaries containing
white dwarfs.  In computing the mass density we have directly summed those stars with known
masses and assigned the mean mass of the local sample to those stars with unknown masses.   The
corresponding mass density of white dwarfs in the local neighborhood is found to be 
3.7$\pm$0.5 x10$^{-3}$M$_{\odot}$  pc$^{-3}$, where the mass uncertainty corresponds to the 
uncertainty in the number density.  This represents only 2\% of the total dynamical mass density 
of 185$\pm$20 x10$^{-3}$ M$_{\odot}$pc$^{-3}$ determined by Bachall (1984).

Our white dwarf density can be compared with other recent published values of the white dwarf space
density ({\em n$_{WD}$}) which have been determined using a variety of methods and which have varied 
by a factor of 2,  ranging from 3.2 to 7.6 x10$^{-3}$pc$^{-3}$.  For example, Sion \& Liebert (1977) 
found 23 white dwarfs within 10 pc and obtained a space density of 5.0 x10$^{-3}$ pc$^{-3}$.  Shipman 
(1983) considering white dwarfs in astrometric binary systems determined n$_{WD}$ = 4.6x10$^{-3}$
pc$^{-3}$ while Liebert, Dahn \& Monet (1988), using a 1/V$_{max}$ method, found local space density 
for single white dwarfs to be 3.2 x10$^{-3}$ pc$^{-3}$. However, Weideman (1991) who also considered 
the number of white dwarfs within 10 pc, revised the Liebert et al. results to suggest a higher value 
of n$_{WD}$ =  5.0 x10$^{-3}$ pc$^{-3}$. More recently, Oswalt et al. (1996) using a 1/V$_{max}$ 
method and estimating the completeness of their sample have estimated a total density of  
n$_{WD}$ = 7.6$^{+3.7}_{-0.7}$x10$^{-3}$pc$^{-3}$ based on observations of a large number of 
wide binaries containing white dwarfs. Leggett, Ruiz \& Bergeron (1998) also using a 1/V$_{max}$ 
technique find  3.39x10$^{-3}$ pc$^{-3}$ and Knox, Hawkins, \& Hambly (1999), using a southern 
hemisphere multi-color proper motion survey, estimate n$_{wd}$=4.16x10$^{-3}$ pc$^{-3}$ from their 
``best guess sample''.  Finally, TT99 considering a number of white dwarfs within 
15 pc find a value of n$_{WD}$  =  4.8x10$^{-3}$pc$^{-3}$. Their lower value of the space density is 
primarily a result of their estimate of 15pc for the completeness distance of the local sample.  
These authors compared the expected and observed numbers of white dwarfs at intervals of 5 pc and 
selected 15 pc as the completeness distance.  If we use a value of 15 pc as our completeness distance 
then we would obtain n$_{WD}$ = 4.4 x10$^{-3}$ pc$^{-3}$ as a density of white dwarfs.

The local sample possesses several advantages over previous estimates of the white dwarf density
derived from other samples.  The chief advantages are its completely volume-limited nature and
relatively high level of apparent completeness.  It does, however, suffer at present from a rather low
sample size.  This limitation, however, is likely to be diminished if the present zone of completeness
is increased from 13 pc to 20 pc.  We intend to do this part of the NSTARS program which is aimed at
discovering and cataloging the stellar population within 20 pc of the Sun.  The primary statistical
uncertainty in all present estimates of the space density of white dwarfs comes from the small sample 
size of typically $\sim$50 stars.  Wood \& Oswalt (1998) used Monte Carlo calculations to estimate that 
for sample sizes of N $\sim$ 50 white dwarfs, uncertainties in n$_{WD}$ of $\sim$50\% are expected.  
There also exists a modest systematic uncertainty due to the possibility that future searches and 
surveys may lead to the discovery of additional white dwarfs within 13 pc.  The slight north-south 
asymmetry in the number of stars within 13 pc, hints at this possibility.  Thus, while it is possible 
that the future may bring a modest increase in the number of white dwarfs in this volume, it is highly 
unlikely, given the quality of the present  distance estimates, that the number of known stars will 
significantly decrease.  Thus, our number and mass density estimates can be regarded as firm lower 
limits.

\subsection{The Composition of the Local Sample}
The types of white dwarf stars which make up the local sample are also of  interest.  This is in part
due to the fact that the population ratio of the two primary spectral types, the H-rich DA stars and
the non-DA stars appears to undergo several changes as a function of effective temperature.  That is, 
white dwarfs appear to change spectral classification based on the dominant atmospheric species as they
cool.   One of the most obvious manifestations of this  is the decline in the DA to non-DA ratio.  At
temperatures of 20,000 K and above, this ratio reaches a value of 7:1 but declines to about 1:1
and lower for white dwarfs near 5,000 K to 4,000 K (BRL).  In the local sample we find the DA to
non-DA ratio to be 1.2. This is consistent with the fact that there are few white dwarfs with 
T$_{eff}$ $>$20,000 K in the local sample.  In this respect, the local sample is quite similar to 
the population of cool white dwarfs studied by BRL, Leggett et al. (1998) and 
Oswalt et al. (1996).  Indeed many stars in these studies, which are drawn heavily from the 
LHS Catalogue (Luyten 1976), are also present in the local sample. It is also similar to the results 
of Sion \& Oswalt (1988) who found a  DA to non-DA ratio of 1 across all spectral types and to that 
found for the Silvestri et al. (2001) sample.  As indicated by Table 2, approximately 30\% of the local 
sample of white dwarfs are members of binary systems.  This includes 29 systems consisting of a white 
dwarf and one or more main sequence stars.  In most instances the main sequence star is an M dwarf.
The above estimate of the binary fraction follows directly from the number of known binaries in our 
sample and thus is, in effect, a lower limit to the true binary fraction.  No attempt has been made 
to correct this estimate for the existence of possible unrecognized binary companions having low masses 
or small separations.  In Table 2 we categorize the local sample separately by spectral type and binary 
nature.  The spectral types in Table 2 are those described in Sion et al. (1983) and used in  MS99.
  
\section{Conclusions}
Using the new version of the Catalog of Spectroscopically Identified White Dwarfs (MS99) we
have identified 118 white dwarfs  having trigonometric and color-based photometric distances within
20 pc of the Sun. The mean mass of this local sample, from published spectroscopic determinations,
is 0.64 M$_{\odot}$.  The local sample of white dwarfs appears to be complete out to a distance of 
13 pc. There are 51 white dwarfs within this completeness radius from which we obtain a mean space
density for white dwarfs in the vicinity of the Sun of 5.5$\pm$0.8x 10$^{-3}$ pc$^{-3}$.  The 
corresponding  mean mass density is found to be 3.7$\pm$0.5 x10$^{-3}$ M$_{\odot}$  pc$^{-3}$.  
The DA to non-DA ratio of this sample is 1.2 and 30\% of the white dwarfs are in binary systems, 
including three double degenerate systems.
%

\begin{deluxetable} {lcl} 
\tablenum{2}
\tablewidth{42pc}
\tablecaption{The Local Population of White Dwarfs}
\vspace{3mm}
\tablehead{
\colhead{Type}    &
\colhead{Number}  &
\colhead{Comments           }}
\startdata
DA &       64 &            H-Rich \nl
DB &        1 &            He-Rich \nl
DC &       33 &            Continuous Spectra - no features \nl
DQ &       13 &            Carbon Features \nl
DZ &        8 &            Metal Lines Only \nl
\tableline
WD + MS &  29 &            White Dwarfs + Main Sequence Binary Systems \nl
WD + WD &   3 &            Double Degenerate Systems \nl
\enddata
\end{deluxetable}

We wish to acknowledge support from NASA grant NAGW5-9408.  Also TDO wishes to thank the NSF for an 
internal grant which partially supported his contribution to this project.

\clearpage

\end{document}